\documentclass[twocolumn,showpacs,preprintnumbers,prb,aps]{revtex4}
\usepackage{amsmath}
\usepackage{color}
\usepackage{graphicx}
\usepackage{dcolumn}

\begin{document}
\title{Observation of a Phase Transition at 55~K in Single-Crystal ${\rm\bf CaCu_{1.7}As_2}$}
\author{V. K. Anand}
\altaffiliation{vanand@ameslab.gov}
\author{D. C. Johnston}
\altaffiliation{johnston@ameslab.gov}
\affiliation {Ames Laboratory and Department of Physics and Astronomy, Iowa State University, Ames, Iowa 50011}

\date{\today}

\begin{abstract}

We present the structural, magnetic, thermal and $ab$-plane electronic transport properties of single crystals of ${\rm CaCu_{1.7}As_2}$ grown by the self-flux technique that were investigated by powder x-ray diffraction, magnetic susceptibility $\chi$, isothermal magnetization $M$, specific heat $C_{\rm p}$, and electrical resistivity $\rho$ measurements as a function of temperature $T$ and magnetic field $H$\@. \mbox{X-ray} diffraction analysis of crushed crystals at room temperature confirm the collapsed  tetragonal ${\rm ThCr_2Si_2}$-type structure with $\sim15$\% vacancies on the Cu sites as previously reported, corresponding to the composition ${\rm CaCu_{1.7}As_2}$.  The $\chi(T)$ data are diamagnetic, anisotropic and nearly independent of $T$\@. The $\chi$  is larger in the $ab$-plane than along the $c$-axis, as also observed previously for ${\rm SrCu_2As_2}$ and for pure and doped BaFe$_2$As$_2$.  The $C_{\rm p}(T)$ and $\rho (T)$ data indicate metallic $sp$-band character.  In contrast to the $\chi(T)$ and $C_{\rm p}(T)$ data that do not show any evidence for phase transitions below 300~K, the $\rho(T)$ data exhibit a sharp decrease on cooling below a temperature $T_{\rm t} =  54$--56~K, depending on the crystal.  The $\rho(T)$ data show no hysteresis on warming and cooling through $T_{\rm t}$ and the transition thus appears to be second order.  The phase transition may arise from spatial ordering of the vacancies on the Cu sublattice.  The $T_{\rm t}$ is found to be independent of $H$ for $H\leq 8$~T\@.  A positive magnetoresistance is observed below $T_{\rm t}$ that increases with decreasing $T$ and attains a value in $H=8.0$~T of 8.7\% at $T=1.8$~K\@.

\end{abstract}


\maketitle

\section{\label{Intro} Introduction}

The discoveries of high-$T_{\rm c}$ superconductivity in iron pnictides and chalcogenides motivated many efforts to identify the mechanism\cite{Rotter2008a, Chen2008, Sasmal2008, Wu2008, Jeevan2008, Sefat2008, Torikachvili2008, Ishida2009, Alireza2009, Johnston2010, Canfield2010, Mandrus2010} and the relationships of these materials to the high-$T_{\rm c}$ cuprates.\cite{Johnston2010, Canfield2010, Mandrus2010, Johnston1997, Damascelli2003, Lee2006} In the latter compounds the copper has a Cu$^{+2}$ $3d^9$ electronic configuration and carries a local magnetic moment with spin $S = 1/2$ which is retained even in the superconducting regime of the phase diagram.  Here we are concerned with the so-called 122-type subclass of the iron arsenide superconductors with the body-centered tetragonal (bct) ${\rm ThCr_2Si_2}$ structure with space group $I4/mmm$.  If a 122-type Cu-arsenide compound having localized $S=1/2$ Cu$^{+2}$ moments could be synthesized, such a compound would bridge the gap between the iron arsenide and cuprate families of high-$T_c$ superconductors. With this in mind, we previously reported the physical properties of ${\rm SrCu_2As_2}$ and ${\rm SrCu_2Sb_2}$ that instead turned out to be nonmagnetic $sp$-band metals,\cite{Anand2012} consistent with the theoretical prediction for SrCu$_2$As$_2$ by Singh.\cite{Singh2009}  His electronic structure calculations for ${\rm BaCu_2As_2}$ and ${\rm SrCu_2As_2}$ indicated that the Cu 3$d$ bands in these compounds are narrow and lie about 3~eV below the Fermi energy $E_{\rm F}$ and therefore the Cu~3$d$ orbitals give very small contributions to the density of states at $E_{\rm F}$.\cite{Singh2009}  From a systematic study of the interlayer $X$--$X$ distance $d_{X-X}$ ($A$ = Ca, Sr, Ba; $X$ = P, As) and the $c/a$ ratio for $AM_2X_2$ $3d$ transition metal $M$ compounds with the ${\rm ThCr_2Si_2}$ structure we concluded that ${\rm SrCu_2As_2}$ crystallizes in a collapsed tetragonal (cT) structure.\cite{Anand2012}

Recent activity in the Fe-based superconductor field has focused on the remarkable physical properties of the class of $A_{1-x}$Fe$_{2-y}$Se$_2$ ($A$ = alkali metal) compounds that are similar to the layered ${\rm ThCr_2Si_2}$-type materials except with substantial numbers of vacancies on the $A$ and Fe sublattices that can become spatially ordered.\cite{Guo2010,Reviews} Depending on the Fe vacancy concentration, superconductivity at temperatures up to $\sim 30$~K and/or large-moment antiferromagnetism with very high N\'eel temperatures up to $\sim 600$~K can be stabilized.  Thus, the influence of transition metal site vacancies on the physical properties of such 122-type compounds is of great current interest.

Here we report powder x-ray diffraction, magnetic susceptibility $\chi$, isothermal magnetization $M$, specific heat $C_{\rm p}$, and $ab$-plane electrical resistivity $\rho$ measurements as a function of temperature $T$ and magnetic field $H$ on single crystals of ${\rm CaCu_{1.7}As_2}$ which was previously  reported by Pilchowski and Mewis to form in the bct ${\rm ThCr_2Si_2}$ structure with a large ($\sim 15$\%) concentration of vacancies on the Cu site.\cite{Pilchowski1990}  To our knowledge, there are no previous studies of the physical properties of this compound including possible temperature-induced Cu vacancy ordering transitions.  Like ${\rm SrCu_2As_2}$, we find that ${\rm CaCu_{1.7}As_2}$ exhibits an anisotropic $T$-independent diamagnetic behavior indicating that the Cu atoms have a nonmagnetic $3d^{10}$ Cu$^{+1}$ electronic configuration.  However, the $\rho(T)$ data for ${\rm CaCu_{1.7}As_2}$ exhibit a phase transition of unknown origin at a transition temperature $T_{\rm t} = 54$--56~K, depending on the crystal, as revealed by a sharp well-defined decrease on cooling below $T_{\rm t}$.  In view of the large ($\sim 15$\%) disordered vacancy concentration on the Cu sites found by Pilchowski and Mewis at room temperature and confirmed by us, this phase transition may reflect the occurrence of Cu vacancy ordering at $T_{\rm t}$.  This possibility could be checked via \mbox{low-$T$} x-ray and/or neutron diffraction measurements.

\section{\label{ExpDetails} Experimental Details}

Single crystals of ${\rm CaCu_{1.7}As_2}$  were grown using prereacted CuAs self-flux starting with the high purity elements Ca (99.98\%), Cu (99.999\%) and As (99.99999\%) from Alfa Aesar. Ca and CuAs were taken in a 1:4 molar ratio and placed in an alumina crucible which was then sealed inside an evacuated quartz tube. The sample was heated to 1100~$^\circ$C at a rate of 60~$^\circ$C/h, held there for 12~h and then cooled to 800~$^\circ$C at a rate of 2.5~$^\circ$C/h at which time the flux was decanted using a centrifuge. Shiny plate-like crystals of typical size $2.5 \times 2 \times 0.3$~mm$^3$ were obtained using this procedure.  All crystals for which measurements are reported here were obtained from the same growth batch.

The structure of the crystals was determined by powder x-ray diffraction (XRD) using Cu K$_\alpha$ radiation on a Rigaku Geigerflex x-ray diffractometer.  The chemical composition was determined by wavelength dispersive x-ray spectroscopy (WDS) analysis using a JEOL JXA-8200 electron probe microanalyzer.

The magnetization $M$ measurements were carried out using a Quantum Design, Inc.\ superconducting quantum interference device-based magnetic properties measurement system in applied magnetic fields $H$ up to 5.5~T\@. A crystal was mounted on a 0.5~mm diameter horizontal rotatable high-purity quartz rod that was inserted in holes in a clear plastic straw attached to the sample hang-down rod.  The sample was attached to the quartz rod using a small amount of GE~7031 varnish.  The contribution to the measured magnetization due to the quartz rod and varnish was measured separately and corrected for.

The $C_{\rm p}(T)$ and $\rho(T,H)$ measurements were carried out using a Quantum Design, Inc.\ physical properties measurement system using the heat capacity and ac transport options, respectively, at fields up to 8~T\@.  The subscript ``p'' in $C_{\rm p}$ refers to measurements at constant pressure.  The $C_{\rm p}$ was measured using a relaxation method.  The $\rho$ was measured in the $ab$-plane using a standard four-probe ac technique with 25~$\mu$m diameter Pt leads attached to the sample with EPO-TEK P1011 silver epoxy that was cured in air at 110~$^\circ$C for 1~h.  We did not cut the as-grown rectangular-shaped crystals \#1 and \#2 that we used for the resistivity measurements because such cutting can potentially  introduce microcracks in the crystals and/or exfoliation of the Cu$_{1.7}$As layers.  The accuracy of the measurements due to uncertainties in the geometric factor is estimated to be $\sim 10$\%.

\section{\label{CaCu2As2} Results and Discussion}

\subsection{Crystallography}

\begin{figure}
\includegraphics[width=3in]{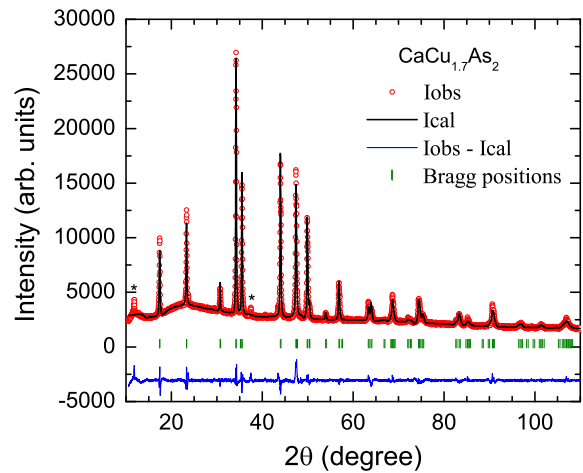}
\caption{(Color online) Powder x-ray diffraction pattern of ${\rm CaCu_{1.7}As_2}$ recorded at room temperature. The solid line through the experimental points is the Rietveld refinement profile calculated for the body-centered tetragonal ThCr$_2$Si$_2$-type structure (space group $I4/mmm$). The short vertical bars mark the fitted Bragg peak positions. The lowermost curve represents the difference between the experimental and calculated intensities. The unindexed peaks marked with stars correspond to peaks from small amounts of flux that could not be removed from the crystals before crushing them for the XRD measurements.}
\label{fig:CaCu2As2_XRD}
\end{figure}
\begin{table}
\caption{\label{tab:XRD1} Crystallographic and Rietveld refinement parameters obtained from powder XRD data of crushed ${\rm CaCu_{1.7}As_2}$ crystals.  Also included are data from Ref.~\onlinecite{Pilchowski1990}.}
\begin{ruledtabular}
\begin{tabular}{lll}
Structure & ${\rm ThCr_2Si_2}$-type\\
Space group & $I4/mmm$ \\
Formula units/unit cell & $Z = 2$\\
\underline{Lattice parameters (RT)} & & \underline{Ref.~\onlinecite{Pilchowski1990}}\\
\hspace{0.8 cm}$a$ (\AA)     &  4.1148(2) &  4.129(1)\\
\hspace{0.8 cm}$c$ (\AA)     &  10.1914(4) & 10.251(1)\\
\hspace{0.8 cm}$c/a$         &  2.4768(2) & 2.482(2)\\
\hspace{0.8 cm}$V_{\rm cell}$ (\AA$^3$) & 172.55(1) & 174.8(1)\\
\underline{Refinement quality} \\
\hspace{0.8 cm}    $\chi^2$	   & 5.63  \\	
 \hspace{0.8 cm}    $R_{\rm p}$ (\%)  & 2.98   \\
  \hspace{0.8 cm}    $R_{\rm wp}$ (\%)  & 4.44  \\
\end{tabular}
\end{ruledtabular}
\end{table}

\begin{table}
\caption{\label{tab:XRD2} Atomic coordinates obtained from the Rietveld refinements of powder XRD data of crushed ${\rm CaCu_{1.7}As_2}$ crystals.  Also included are data from Ref.~\onlinecite{Pilchowski1990}.}
\begin{ruledtabular}
\begin{tabular}{ccccccc}
  Atom & Wyckoff   &	 $x$ 	&	$y$	&	$z$	 & Fractional & Ref.\\	
   	& position 		& 			&		& 		& occupancy\\
   	& 				& 			&		& 		& (\%)\\
\hline
    Ca & 2a  	&	 0 	&	0	&   0		& 100  & This work\\
    Ca & 2a  	&	 0 	&	0	&   0		& 100  & \onlinecite{Pilchowski1990}\\
    Cu & 4d	    &	 0 	&	1/2	&	1/4 	& 82.5(5)  & This work\\
    Cu & 4d	    &	 0 	&	1/2	&	1/4 	& 87.5(8)  & \onlinecite{Pilchowski1990}\\
    As & 4e 	&	 0 	&   0 	&	0.3779(2) & 100 & This work\\
    As & 4e 	&	 0 	&   0 	&	0.3799(2) & 100 & \onlinecite{Pilchowski1990}
\end{tabular}
\end{ruledtabular}
\end{table}

Powder XRD data were collected on crushed ${\rm CaCu_{1.7}As_2}$ single crystals at room temperature (RT) and analyzed by Rietveld refinement using the {\tt FullProf} software. \cite{Rodriguez1993} Figure~\ref{fig:CaCu2As2_XRD} shows the XRD data and the \mbox{Rietveld} fit profile. The refinement confirmed the single phase nature of the crystals and the ${\rm ThCr_2Si_2}$-type body-centered tetragonal structure with space group $I4/mmm$ found previously by Pilchowski and Mewis from a single-crystal structure refinement.\cite{Pilchowski1990}  Our crystallographic and refinement parameters are listed in Tables~\ref{tab:XRD1} and \ref{tab:XRD2} and compared with those of Pilchowski and Mewis.\cite{Pilchowski1990} The lattice parameters $a$ and $c$ and the $z$-coordinate of the As atoms $z_{\rm As}$ that we obtained are in good agreement with the literature values.\cite{Pilchowski1990}

While refining the XRD powder pattern for ${\rm CaCu_{1.7}As_2}$ we noticed that the lattice parameters and the As $c$-axis position parameter $z_{\rm As}$ were insensitive to changes in the thermal parameters $B$ within the error bars, so we kept $B$ fixed to $B \equiv 0$.  On the other hand, the refinement quality and the calculated line intensities were very sensitive to the fractional occupancy of the 4d sites by Cu.  Therefore we allowed the occupancy of this site by Cu to vary during the refinement, while the occupancies of the Ca and As positions were kept fixed at the stoichiometric values of unity.  As shown in Table~\ref{tab:XRD2}, the refined values of the Cu site occupancy obtained both by us and by Pilchowski and Mewis show a large vacancy concentration on the Cu sites of $\approx 15$\%, corresponding to an approximate composition of ${\rm CaCu_{1.7}As_2}$.  However, there are no indications of superstructure reflections in the XRD patterns that would indicate ordering of the Cu vacancies at room temperature and Pilchowski and Mewis also did not report such reflections.  Thus we assume that the Cu vacancies are randomly distributed on the Cu sites at room temperature.

Our WDS analysis at about ten points on the $ab$ plane of a ${\rm CaCu_{1.7}As_2}$ crystal gave the average atomic ratios as Ca\,:\,Cu\,:\,As = 21.4(2)\,:\,36.5(3)\,:\,42.1(2), corresponding to the stoichiomentry ${\rm Ca_{1.02(2)}Cu_{1.73(2)}As_2}$ assuming the occupancy of the As site to be 100\%.  This result confirms complete occupancy of the Ca site and again indicates a large vacancy concentration on the Cu sites.

For the molar heat capacity, magnetization and magnetic susceptibility data presented below, a ``mole'' is defined as a mole of ${\rm CaCu_{1.7}As_2}$ formula units (f.u.).

From the crystal data in Tables~\ref{tab:XRD1} and \ref{tab:XRD2} we obtain the interlayer As--As distance as $d_{\rm As-As} = (1-2z_{\rm As})c = 2.49$~\AA. This value of $d_{\rm As-As}$ for ${\rm CaCu_{1.7}As_2}$ is close to the covalent (single) bond distance 2.38~\AA\ for As.\cite{Cordero2008}  Furthermore, the values of $c/a$ and $d_{\rm As-As}$ fall in the respective ranges for the collapsed tetragonal structure compounds shown in Fig.~22 of Ref.~\onlinecite{Anand2012}. This shows that like ${\rm SrCu_2As_2}$, ${\rm CaCu_{1.7}As_2}$ also has a cT structure.  A consequence of the formation of the cT structure is an unusual oxidation state of As$^{-2} \equiv $ [As--As]$^{-4}$, which together with Ca in the Ca$^{+2}$ oxidation state, indicates that the Cu in ${\rm CaCu_{1.7}As_2}$ has an oxidation state of $\approx +1.2$. However, the nonmagnetic nature of this compound deduced from our magnetic measurements presented below in Sec.~\ref{Sec:CaCu2As2_ChiMH} suggests instead a filled Cu $3d$ shell, a Cu electronic configuration $3d^{10}$, and formal oxidation states Cu$^{+1}$ and As$^{-1.85}$.  The latter value suggests the presence of hole conduction on the As sublattice.

\subsection{\label{Sec:CaCu2As2_HC} Heat Capacity}

\begin{figure}
\includegraphics[width=3in]{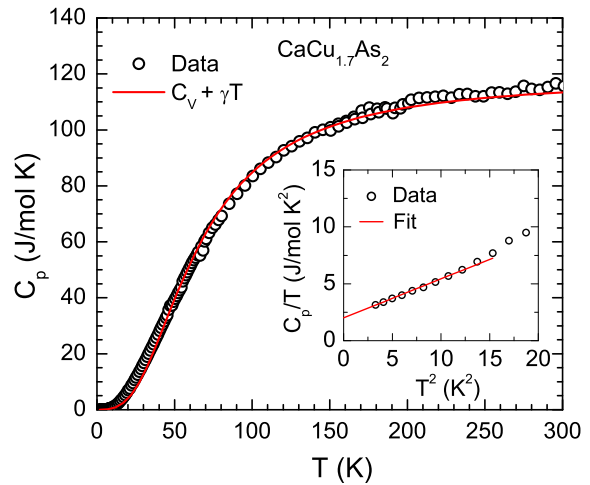}
\caption{(Color online) Heat capacity $C_{\rm p}$ of a ${\rm CaCu_{1.7}As_2}$ single crystal as a function of temperature $T$ measured in zero magnetic field. The red solid curve is the fitted sum of the contributions from the Debye lattice heat capacity $C_{\rm V\,Debye}(T)$ and predetermined electronic heat capacity $\gamma T$ according to Eq.~(\ref{eq:Debye_HC-fit}). Inset: $C_{\rm p}/T$ versus $T^2$ for $T \leq 4.5$~K\@. The straight red line is a fit of the data by $C_{\rm p}/T = \gamma + \beta T^2$ for $1.8~{\rm K} \leq T \leq 3.5$~K\@.}
\label{fig:HC_CaCu2As2}
\end{figure}

Figure~\ref{fig:HC_CaCu2As2} shows $C_{\rm p}$ versus $T$ of a ${\rm CaCu_{1.7}As_2}$ crystal from 1.8 to 300~K\@.  No obvious anomaly that might be associated with the occurrence of a phase transition is observed over this $T$ range.  The value of $C_{\rm p}(300$~K) is $\approx 116$~J/mol\,K, which is close to the classical Dulong-Petit prediction of the lattice heat capacity $C_{\rm V} = 3nR = 14.1R$ = 117.2~J/mol\,K at constant volume, where $n$ is the number of atoms per formula unit ($n=4.7$ here) and $R$ is the molar gas constant. \cite{Kittel2005, Gopal1966}

A conventional linear fit of $C_{\rm p}(T)/T = \gamma + \beta T^2$ in the temperature range $1.8~{\rm K} \leq T\leq 3.5$~K is shown in the inset of Fig.~\ref{fig:HC_CaCu2As2} and gives the electronic Sommerfeld specific heat coefficent $\gamma=2.0(2)$~mJ/mol\,K$^2$ and the lattice heat capacity coefficient $\beta= 0.33(2)$~mJ/mol\,K$^4$. The density of states at the Fermi energy ${\cal D}(E_{\rm F})$ for both spin directions is estimated from $\gamma$  using the single-band relation \cite{Kittel2005} $\gamma = (\pi^2 k_{\rm B}^2/3) {\cal D}(E_{\rm F})$, from which we obtain ${\cal D}(E_{\rm F}) = 0.85(9)$~states/eV\,f.u.\ for both spin directions. The Debye temperature $\Theta_{\rm D}$ is estimated from $\beta$ using the relation \cite{Kittel2005} $\Theta_{\rm D} = (12 \pi^{4} n R/5 \beta)^{1/3}$, giving $\Theta_{\rm D}= 303(6)$~K\@.

The entire $C_{\rm p}(T)$ data set from 1.8 to 300~K was fitted by
\begin{equation}
C_{\rm p}(T) = \gamma T + n C_{\rm{V\,Debye}}(T),
\label{eq:Debye_HC-fit}
\end{equation}
where $\gamma$ was fixed to the value $\gamma = 2.0~{\rm mJ/mol\,K^2}$ obtained above and the Debye heat capacity $C_{\rm{V\,Debye}}(T)$ describes the heat capacity due to acoustic phonons at constant volume~V and is given by \cite{Gopal1966}
\begin{equation}
C_{\rm{V\,Debye}}(T) = 9 R \left( \frac{T}{\Theta_{\rm{D}}} \right)^3 {\int_0^{\Theta_{\rm{D}}/T} \frac{x^4 e^x}{(e^x-1)^2}\,dx}.
\label{eq:Debye_HC}
\end{equation}
The fit is shown as the solid red curve in Fig.~\ref{fig:HC_CaCu2As2}.  In the fit, we used our analytic Pad\'e approximant function given in Ref.~\onlinecite{Ryan2012} that accurately represents $C_{\rm{V\,Debye}}(T)$ and obtained $\Theta_{\rm D} = 265(1)$~K, which is somewhat smaller than the value $\Theta_{\rm D} = 303(6)$~K obtained from fitting the \mbox{low-$T$} $C_{\rm p}(T)$ data above.  The difference between these two values reflects the $T$ dependence of $\Theta_{\rm D}$.\cite{Gopal1966, Ryan2012}

\subsection{\label{Sec:CaCu2As2_ChiMH} Magnetization and Magnetic Susceptibility}

\begin{figure}
\includegraphics[width=3in]{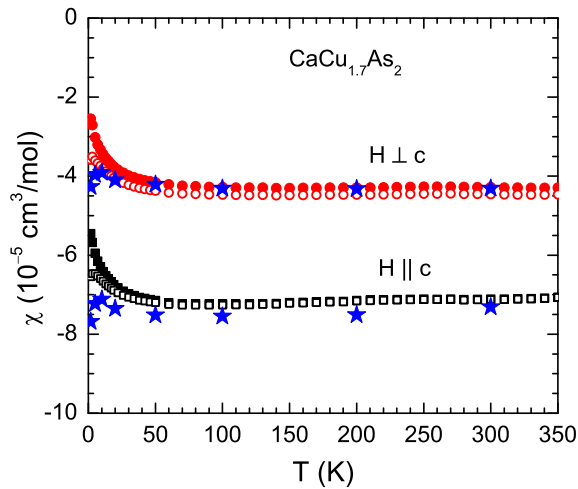}
\caption{(Color online) Zero-field-cooled magnetic susceptibility $\chi$ of a ${\rm CaCu_{1.7}As_2}$ single crystal versus temperature $T$ in the temperature range 1.8--350~K measured in a magnetic field $H = 3.0$~T applied along the $c$-axis ($\chi_c,\ H \parallel {\bf c}$) and in the $ab$-plane ($\chi_{ab},\ H \perp  {\bf c}$) (solid symbols). The open symbols represent the intrinsic susceptibility of ${\rm CaCu_{1.7}As_2}$ after correcting for the ferromagnetic and paramagnetic impurity contributions as described in the Appendix. The filled blue stars represent the intrinsic $\chi$ obtained from fitting $M(H)$ isotherm data by Eq.~(\ref{eq:MH_fit}) and are more reliable.}
\label{fig:MT_CaCu2As2}
\end{figure}

Figure~\ref{fig:MT_CaCu2As2} shows the zero-field-cooled magnetic susceptibility $\chi \equiv M/H$ of a ${\rm CaCu_{1.7}As_2}$  crystal as a function of $T$ from 1.8 to 350~K for $H = 3.0$~T applied along the $c$-axis ($\chi_c,\ H \parallel c$) and in the $ab$-plane ($\chi_{ab},\ H \perp c$). The $\chi(T)$ data for both directions of $H$ are diamagnetic and nearly independent of $T$\@. The $\chi_{c}$ is significantly more negative than $\chi_{ab}$.  The same type of $\chi$ anisotropy was previously observed for ${\rm SrCu_2As_2}$,\cite{Anand2012} BaFe$_2$As$_2$,\cite{Wang2009} SrFe$_2$As$_2$,\cite{Yan2008} and other doped and undoped FeAs-based compounds.\cite{Johnston2010}

Curie-like upturns occur in $\chi(T)$ at low~$T$ in Fig.~\ref{fig:MT_CaCu2As2} that are likely due at least in part to the presence of small amounts of saturable paramagnetic (PM) impurities in the ${\rm CaCu_{1.7}As_2}$ crystal. Our analysis of $M(H)$ isotherms in the Appendix allows us to approximately correct for such contributions. The intrinsic susceptibilities after corrections for paramagnetic and ferromagnetic (FM) impurity contributions are shown as open symbols in Fig.~\ref{fig:MT_CaCu2As2}.  The corrected susceptibilities still show small residual upturns at low~$T$ which, in view of the intrinsic $T$-independent susceptibilities obtained at eight temperatures from analysis of the $M(H)$ isotherms (shown in Fig.~\ref{fig:MT_CaCu2As2} as filled blue stars), are due to inaccuracies in correcting for the impurity contributions to the $M/H$ versus $T$ data.

The intrinsic $\chi$ consists of different contributions given by
\begin{equation}
\chi=\chi_{\rm {core}}+\chi_{\rm {VV}}+\chi_{\rm {L}} + \chi_{\rm {P}}.
\label{eq:chi}
\end{equation}
The first three terms are orbital susceptibilities.  $\chi_{\rm {core}}$ is the diamagnetic core susceptibility, $\chi_{\rm {VV}}$ is the paramagnetic Van Vleck susceptibility and $\chi_{\rm {L}}$ the diamagnetic Landau susceptibility of the conduction electrons.  The last term $\chi_{\rm {P}}$ is the paramagnetic Pauli spin susceptibility. The $\chi_{\rm {core}}$ = $-1.53 \times 10^{-4}$~cm$^3$/mol is estimated using the atomic diamagnetic susceptibilities. \cite{Mendelsohn1970}  $\chi_{\rm {P}}$ is estimated from $\chi_{\rm {P}} =  \mu_{\rm B}^2 {\cal D}(E_{\rm F})$ (Ref.~\onlinecite{Ashcroft1976}), giving $\chi_{\rm {P}} = 2.7(3) \times 10^{-5}$~cm$^3$/mol using ${\cal D}(E_{\rm F}) = 0.85(9)$~states/eV\,f.u.\ for both spin directions obtained above in Sec.~\ref{Sec:CaCu2As2_HC}.  The $\chi_{\rm {L}}$ is obtained from $\chi_{\rm {L}} = - \frac{1}{3} \left( \frac {m_{\rm e}}{m^*} \right)^2 \chi_{\rm {P}}$,\cite{Ashcroft1976, Elliott1998} which gives $\chi_{\rm {L}} = -0.9 \times 10^{-5}$~cm$^3$/mol assuming that the effective mass $m^*$ equals the free electron mass $m_{\rm e}$.  The angle and temperature average of the anisotropic $\chi$ in Fig.~\ref{fig:MT_CaCu2As2} over the $T$ range 30 to 350~K is $\langle\chi\rangle = [2 \langle\chi_{ab}\rangle + \langle\chi_{c}\rangle]/3 = -5.3\times 10^{-5}$~cm$^3$/mol. We can now estimate $\langle\chi_{\rm {VV}}\rangle$ using the above estimated values of $\chi_{\rm {core}}$, $\chi_{\rm {P}}$ and $\chi_{\rm {L}}$ yielding the powder-averaged $\langle\chi_{\rm {VV}}\rangle = 8.2 \times 10^{-5}$~cm$^3$/mol from Eq.~(\ref{eq:chi}), which is a physically realistic value.  The $T$-independent  anisotropic Van Vleck contributions are $\chi_{\rm {VV}}^c = 6.3 \times 10^{-5}$~cm$^3$/mol and $\chi_{\rm {VV}}^{ab} = 9.1 \times 10^{-5}$~cm$^3$/mol for $\ H \parallel c$ and $\ H \perp c$, respectively.

\subsection{\label{Sec:CaCu2As2ChiMH} Electrical Resistivity}
\begin{figure}
\includegraphics[width=3in]{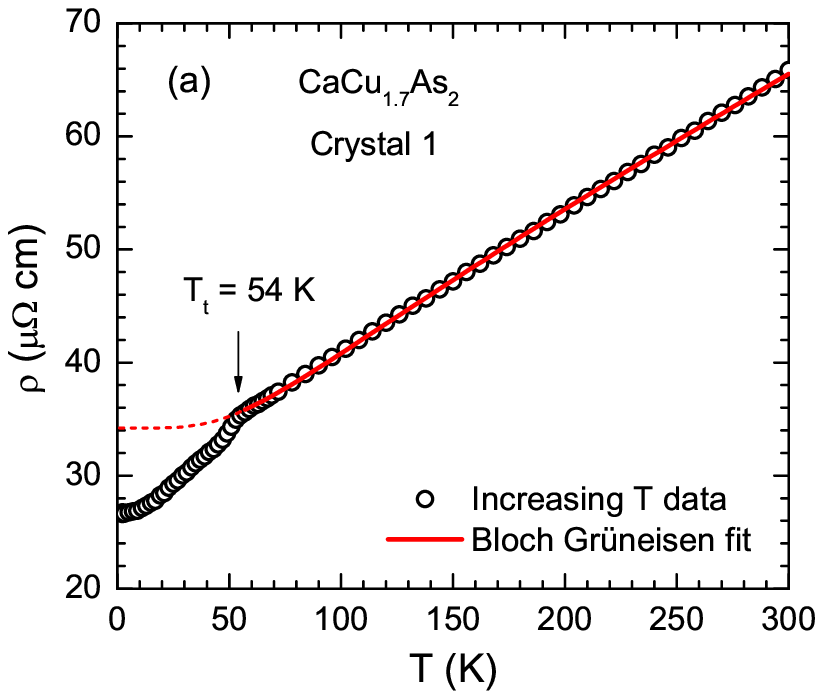}\vspace{0.1in}
\includegraphics[width=3in]{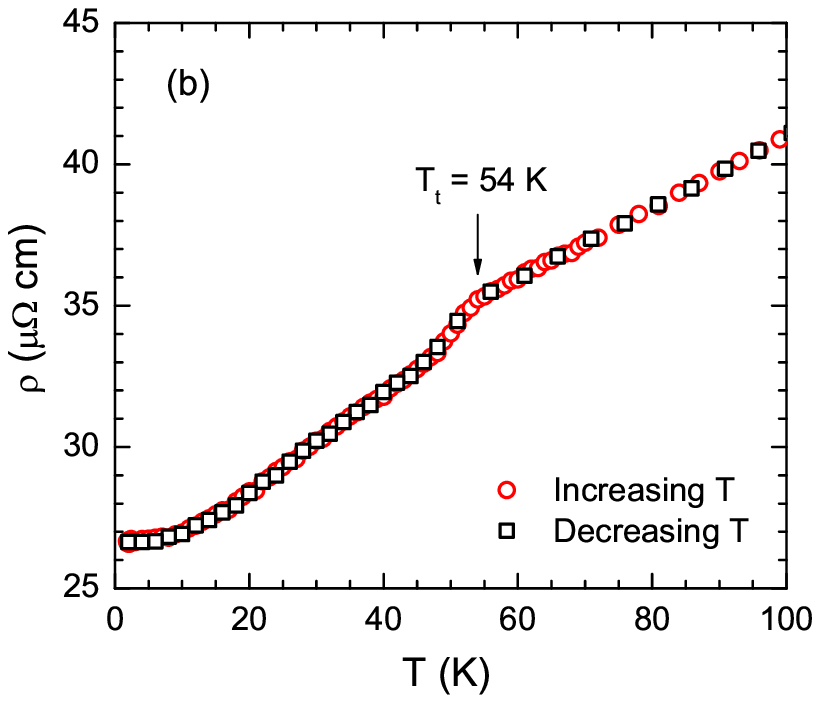}
\caption{(Color online) (a) In-plane electrical resistivity $\rho$ of ${\rm CaCu_{1.7}As_2}$ single crystal \#1 as a function of temperature $T$ measured in zero magnetic field (open circles) showing a  transition at $T_{\rm t} = 54$~K\@. A fit by the Bloch-Gr\"uneisen model from 57 to 300~K is shown by the solid red curve, and an extrapolation of the fit to $T=0$ is shown as the dashed red curve.  (b) Expanded plot of $\rho$ versus $T$ for heating and cooling cycles through $T_{\rm t}$ at $T<100$~K\@.}
\label{fig:rho_CaCu2As2}
\end{figure}

\begin{figure}
\includegraphics[width=3in]{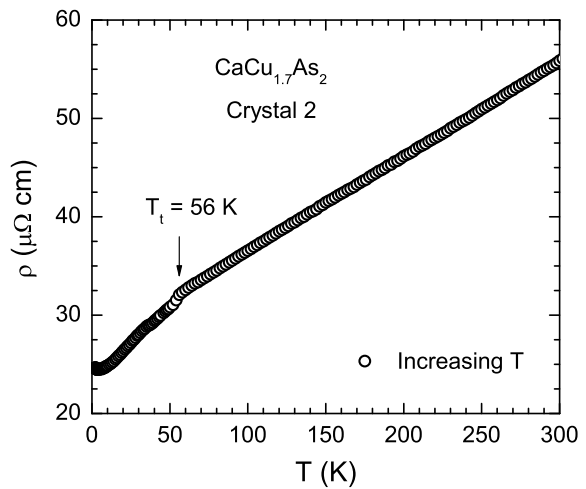}
\caption{ In-plane electrical resistivity $\rho$ of ${\rm CaCu_{1.7}As_2}$ single crystal \#2 versus temperature $T$ in zero magnetic field showing the same shape of transition at $T_{\rm t} = 56$~K as seen for crystal~\#1 at 54~K in Fig.~\ref{fig:rho_CaCu2As2}. }
\label{fig:rho_CaCu2As2No2}
\end{figure}

Figure~\ref{fig:rho_CaCu2As2} shows $\rho(T)$ of ${\rm CaCu_{1.7}As_2}$ crystal~\#1 from 1.8 to 300~K in $H=0$. The residual resistivity at $T= 1.8$~K is $\rho_0 = 26.7(1)\,\mu\Omega\,{\rm cm}$ and the residual resistivity ratio is ${\rm RRR} \equiv \rho(300\,{\rm K})/\rho_0 \approx 2.5$. The $\rho(T)$ data exhibit metallic behavior with an almost linear $T$ dependence of $\rho$ above 55~K\@.  A sharp decrease is observed in $\rho(T)$ on cooling below a transition temperature $T_{\rm t} = 54$~K\@.  This behavior is reproduced without hysteresis upon heating and cooling through $T_{\rm t}$ as shown in Fig.~\ref{fig:rho_CaCu2As2}(b), suggesting a second-order transition.  The transition anomaly at 54~K was reproduced in a $\rho(T)$ measurement on another crystal~\#2 for which we found $T_{\rm t} = 56$~K, as shown in Fig.~\ref{fig:rho_CaCu2As2No2}.  As noted in previous sections, no evidence for the transition was observed in our $\chi(T)$ measurements, suggesting that the transition is not magnetic in nature.  The transition may be associated with spatial ordering of the Cu vacancies discussed above.

\begin{figure}
\includegraphics[width=3in]{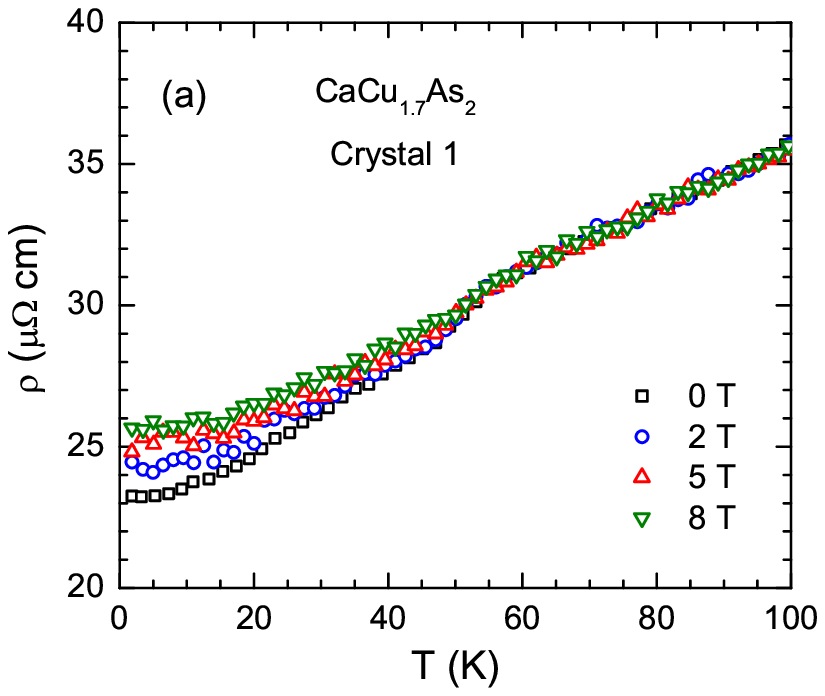}\vspace{0.1in}
\includegraphics[width=3in]{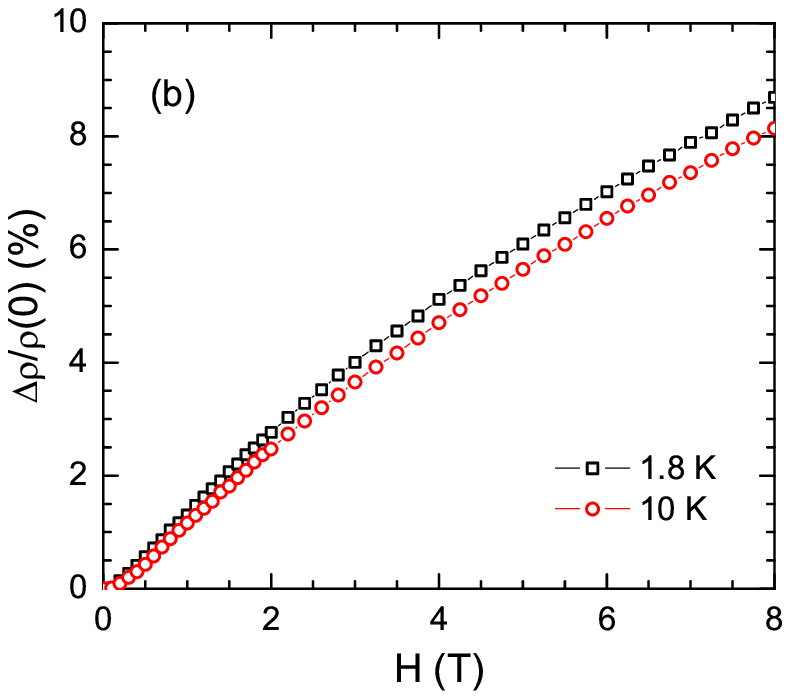}
\caption{(Color online) (a) In-plane electrical resistivity $\rho$ of a ${\rm CaCu_{1.7}As_2}$ single crystal versus temperature $T$ measured in the indicated magnetic fields $H$. The $\rho(T)$ data in this figure were measured after remounting the leads, resulting in a slight change in the absolute values of $\rho$ between this figure and Fig.~\ref{fig:rho_CaCu2As2}.  (b) Magnetoresistance $[\rho(H)-\rho(0)]/\rho(0)$ versus $H$ at $T=1.8$ and 10~K\@.}
\label{fig:rho_CaCu2As2_H}
\end{figure}

We also measured $\rho(T < 100)$~K in high magnetic fields $H\leq 8.0$~T as shown in Fig.~\ref{fig:rho_CaCu2As2_H}(a). The transition temperature $T_{\rm t}$ is found to be independent of $H$ over this field range.  However, a magnetoresistance (MR) $\Delta \rho/\rho(0) = [\rho(H)-\rho(0)]/\rho(0)$  is observed at temperatures $T<T_{\rm t}$ and is plotted versus $H$ for $T =  1.8$ and 10~K in Fig.~\ref{fig:rho_CaCu2As2_H}(b).  The MR is positive up to 8.0~T with MR~$\approx$ 8.7\% at 1.8 K and $H = 8.0$~T\@.  The reason that measurable MR values are only observed at $T<T_{\rm t}$ is not clear.

Our detailed $\rho(T,H)$ measurements on crystal~\#1 indicate that the transition at $T_{\rm t} = 54$~K is not due to microcracks in the sample because the data on heating and cooling the same sample in Figs.~\ref{fig:rho_CaCu2As2} and \ref{fig:rho_CaCu2As2_H}(a) reproducibly show the same sharp transition temperature with the same shape of the anomaly, and the transition occurs reproducibly between samples as seen in Fig.~\ref{fig:rho_CaCu2As2No2}.  Furthermore, we can see no mechanism by which the magnetoresistance that is observed only below $T_{\rm t}$ in Fig.~\ref{fig:rho_CaCu2As2_H} could arise from microcracks.

The transition in ${\rm CaCu_{1.7}As_2}$ seen in $\rho(T)$ at 54--56~K in Figs.~\ref{fig:rho_CaCu2As2} and~\ref{fig:rho_CaCu2As2No2} might potentially be due to an extrinsic phase transition in residual Cu-As flux attached to the crystals.  Three binary phases exist in the Cu-As binary phase diagram with compositions of approximately Cu$_3$As, ${\rm Cu_5As_2}$ and Cu$_2$As.  Previous $\rho,\ \chi$ and nuclear quadrupole resonance measurements versus $T$ on these phases\cite{Pauwels1973, Begaev2002} showed no phase transitions at any temperature near our phase transition temperature \mbox{$T_{\rm t} = 54$--56 K\@.}  Moreover, the ${\rm SrCu_2As_2}$ crystals for which we reported the properties in Ref.~\onlinecite{Anand2012} were grown using the same CuAs self-flux, and these crystals showed no evidence for a transition at $T \approx 55$~K in our $\rho(T)$ measurements on them.  Thus we rule out this extrinsic cause of the transition we see in ${\rm CaCu_{1.7}As_2}$ at $T_{\rm t}$.

The lack of an obvious heat capacity anomaly at $\sim 55$~K in Fig.~\ref{fig:HC_CaCu2As2} might be construed as evidence that the second-order phase transition at $T_{\rm t} = 54$--56~K indicated by the resistivity data in Figs.~\ref{fig:rho_CaCu2As2}--\ref{fig:rho_CaCu2As2_H} is not a bulk effect.  However, an observable change in the heat capacity is only expected if the temperature derivative $dS/dT$ of the entropy $S$ changes sufficiently strongly at $T_{\rm t}$.  In particular, the heat capacity of a material is $C_{\rm p} = dQ/dT = T\,dS/dT$, where $dQ$ is the increment of heat absorbed and $dS = dQ/T$ is the incremental change in the total entropy of the system.  Bulk phase transitions can occur that involve very little change in the slope of the entropy versus temperature of the system.  For example, the bulk second-order antiferromagnetic phase transition at the N\'eel temperature $T_{\rm N} \sim 300$~K of the layered cuprate ${\rm La_2CuO_4}$ has not been observed by heat capacity measurements because the change in $dS/dT$ at $T_{\rm N}$ is too small.\cite{Sun1991}  Thus the lack of an observable heat capacity anomaly at $T_{\rm t}$ in Fig.~\ref{fig:HC_CaCu2As2} does not rule out a bulk phase transition at that temperature.

In the Bloch-Gr\"{u}neisen (BG) model, the resistivity arises from scattering of conduction electrons by longitudinal acoustic lattice vibrations, given by\cite{Blatt1968}
\begin{equation}
\rho_{\rm {BG}}(T)= 4 \mathcal{R}(\Theta _{\rm R}) \left( \frac{T}{\Theta _{\rm{R}}}\right)^5 \int_0^{\Theta_{\rm{R}}/T}{\frac{x^5}{(e^x-1)(1-e^{-x})}dx},
 \label{eq:BG}
\end{equation}
where $\Theta_{\rm{R}}$ is the Debye temperature obtained from fitting resistivity data. For polyatomic systems the prefactor $\mathcal{R}(\Theta _{\rm R})$ is given by \cite{Anand2012, Ryan2012}
\begin{equation}
 \mathcal{R}(\Theta _{\rm R})=\frac{\hbar}{e^2} \left[ \frac{\pi^3 (3 \pi^2)^{1/3} \hbar^2}{4 n_{\rm{cell}}^{2/3} a^\ast k_{\rm{B}} \Theta_{\rm{R}}} \left(\frac{1}{M}\right)_{\rm ave} \right]
 \label{eq:BG-R}
\end{equation}
where $\hbar$ is Planck's constant divided by $2\pi$, $k_{\rm B}$ is Boltzmann's constant, $e$ is the fundamental electric charge, $n_{\rm cell}$ is the number of conduction (valence) electrons per atom, $(1/M)_{\rm ave}$ is the average inverse mass of the atoms in the compound, and $a^\ast =[V_{\rm cell}/nZ]^{1/3}$ is the equivalent lattice parameter of a primitive cubic unit cell containing one atom, $Z$ being the number of formula units per unit cell and $n$ the number of atoms per f.u.

As discussed in Refs.~\onlinecite{Ryan2012} and~\onlinecite{Blatt1968}, it is usually not possible to obtain an accurate fit to $\rho(T)$ data by the BG model with the single adjustable parameter $\Theta_{\rm R}$.  Therefore we allowed the prefactor in Eq.~(\ref{eq:BG}) to vary independently and fitted our in-plane $\rho(T>T_{\rm t})$ data for ${\rm CaCu_{1.7}As_2}$ in Fig.~\ref{fig:rho_CaCu2As2}(a) by
\begin{equation}
\rho(T) = \rho_0' + \rho(\Theta_{\rm R}) f(T/\Theta_{\rm R}),
\label{eq:BG_fit}
\end{equation}
where $\rho_0'$ is the residual resistivity extrapolated from $T>T_{\rm t}$ and from Eq.~(\ref{eq:BG}) one obtains\cite{Anand2012, Ryan2012}
\begin{equation}
\begin{split}
f(y) & = \frac{\rho_{\rm BG}(T)}{\rho_{\rm BG}(T=\Theta_{\rm R})} \\
& = 4.226\,259 \,y^5 \int_0^{1/y}\frac{x^5}{(e^x - 1)(1-e^{-x})}\,dx,
\label{eq:BG_fn}
\end{split}
\end{equation}
where $y=T/\Theta_{\rm R}$ and
\begin{equation}
\rho_{\rm BG}(T=\Theta_{\rm R})=0.9\,464\,635\,{\cal R}(\Theta _{\rm R}).
\label{eq:BG_R}
\end{equation}
A fit of $\rho(T)$ data by Eqs.~(\ref{eq:BG_fit}) and (\ref{eq:BG_fn}) thus has three independent adjustable parameters $\rho_0'$, $\rho(\Theta_{\rm R})$ and $\Theta_{\rm R}$.

A good fit of the $\rho(T)$ data in Fig.~\ref{fig:rho_CaCu2As2}(a) for 57~K~$\leq T \leq$~300~K was obtained, as shown by the solid red curve in Fig.~\ref{fig:rho_CaCu2As2}(a) where we used an accurate analytic Pad\'e approximant function of $y$ in place of Eq.~(\ref{eq:BG_fn}) as given in Ref.~\onlinecite{Ryan2012}.  The parameters obtained from the fit are  $\rho_0' =34.2(1)\,\mu\Omega$\,cm, $\rho(\Theta_{\rm{R}}) = 33.7(6)\,\mu\Omega$\,cm and $\Theta_{\rm{R}} = 320(6)$~K\@. The $\mathcal{R}(\Theta _{\rm R})$ calculated from the value of $\rho(\Theta_{\rm{R}})$ using Eq.~(\ref{eq:BG_R}) is $\mathcal{R}(\Theta _{\rm R}) = 35.6\,\mu\Omega$\,cm.  In order to compare the resistivity at, e.g. room temperature, with the value predicted by the BG theory, one needs an estimate of the conduction carrier concentration $n_{\rm cell}$ in Eq.~(\ref{eq:BG-R}).  Such an estimate is not currently available.

\section{\label{Conclusion} Conclusions}

\begin{table}
\caption{\label{Tab:Parameters} Values of parameters obtained from analyses of heat capacity, magnetic susceptibility and electrical resistivity measurements of ${\rm CaCu_{1.7}As_2}$.  The notation $\langle\cdots\rangle$ denotes a temperature and/or angular average of the enclosed quantity.}
\begin{ruledtabular}
\begin{tabular}{ll}
 Property  & Value \\
\hline
\underline{Heat Capacity}\\
$\gamma$ & 2.0(2)~mJ/mol\,K$^2$ \\
$\beta$  & 0.33(2)~mJ/mol\,K$^4$ \\
${\cal D}(E_{\rm F})$ & 0.85(9)~states/eV\,f.u.\ (both spin directions) \\
$\Theta_{\rm D} $ & 303(6)~K (from low-$T$)\\
$\Theta_{\rm D} $ & 265(1)~K (from all $T$) \\				
\\
\underline{Susceptibility} \\
$\langle \chi \rangle$ & $-5.3\times 10^{-5}$~cm$^3$/mol\\
$\chi_{\rm core}$ & $-1.53 \times 10^{-4}$~cm$^3$/mol \\
$\chi_{\rm {P}}$ &	$2.7(3) \times 10^{-5}$~cm$^3$/mol \\
$\chi_{\rm L}$	& $-0.9 \times 10^{-5}$~cm$^3$/mol	\\
$\langle\chi_{\rm VV}\rangle$ & $8.2 \times 10^{-5}$~cm$^3$/mol \\
$\chi_{\rm {VV}}^{ab}$ & $9.1 \times 10^{-5}$~cm$^3$/mol\\
$\chi_{\rm {VV}}^{c}$ & $6.3 \times 10^{-5}$~cm$^3$/mol\\
\\
\underline{Resistivity}\\
$\rho_0$ & 26.7(1)~$\mu \Omega\,{\rm cm}$ \\
RRR &  $\approx 2.5$ \\
$\rho_0'$ & 34.2(1)~$\mu \Omega$\,cm \\
$\rho(\Theta_{\rm{R}})$ & 33.7(6)~$\mu \Omega$\,cm \\
$\mathcal{R}(\Theta _{\rm R}) $  & 35.6~$\mu \Omega$\,cm\\ 	
$\Theta_{\rm{R}}$ & 320(6)~K \\
\end{tabular}
\end{ruledtabular}
\end{table}

We have successfully grown single crystals of ${\rm CaCu_{1.7}As_2}$ and investigated their crystallographic, magnetic, thermal, and electronic transport properties. \mbox{Rietveld} refinements of powder XRD data for crushed crystals  and WDS chemical analyses of single crystal surfaces indicate the presence of $\approx 15$\% vacancies on the Cu sites, consistent with literature data.  No superconductivity was observed above 1.8~K\@. Our crystallographic and refinement parameters are listed in Tables~\ref{tab:XRD1} and \ref{tab:XRD2} and a summary of the parameters obtained from our various physical property measurements on ${\rm CaCu_{1.7}As_2}$ is given in Table~\ref{Tab:Parameters}.  The $\chi(T)$ data reveal a nearly $T$-independent anisotropic diamagnetic behavior, indicating that the Cu atoms in ${\rm CaCu_{1.7}As_2}$ are in the Cu$^{+1}$ oxidation state with a nonmagnetic $3d^{10}$ electronic configuration, as expected from the collapsed-tetragonal crystal structure of this compound. The formal oxidation state of the As, which participates in As--As interlayer bonding, is then As$^{-1.85}$ which suggests hole conduction on the As sublattice.  The $C_{\rm p}(T)$ and $\rho (T)$ data reveal metallic behavior. A small density of states at the Fermi level is found, consistent with ${\rm CaCu_{1.7}As_2}$ being an $sp$-band metal. The overall $C_{\rm p}(T)$ and $\rho(T > T_{\rm t})$ behaviors are well-described by the Debye model and the Bloch-Gr\"{u}neisen model, respectively.  However, the $\rho(T)$ of ${\rm CaCu_{1.7}As_2}$ exhibits a transition of unknown origin at $T_{\rm t} = 54$--56~K without any thermal hysteresis, suggesting that the transition is  second order.  A significant positive magnetoresistence develops below this transition temperature.  This transition may arise from spatial ordering of the Cu vacancies on cooling below $T_{\rm t}$.  High-resolution x-ray and/or neutron diffraction measurements at low~$T$ could test this hypothesis.

\noindent\emph{Note Added ---} After submission of this paper, \mbox{Cheng~{\it et al.}}\cite{Cheng2012} reported in this journal the observation of a transition in ``${\rm CaCu_2As_2}$'' single crystals at 50~K from $\rho(T)$ measurements.  As in the present paper, their $\chi(T)$ data for this compound showed no evidence of the transition.    These authors did not mention the large concentration of vacancies on the Cu sites previously reported in Ref.~\onlinecite{Pilchowski1990} and  confirmed by us.  Cheng~{\it et al.} noted that the shape of the transition in $\rho(T)$ is similar to those observed for CaFe$_2$(As$_{1-x}$P$_x)_2$ (Ref.~\onlinecite{Kasahara2011}) and Ca$_{1-x}R_x$Fe$_2$As$_2$ ($R$ = lanthanide, Ref.~\onlinecite{Saha2012}) arising from transitions from tetragonal to collapsed-tetragonal (cT) structures on cooling below room temperature.  However, as we have discussed herein and previously,\cite{Anand2012} ${\rm CaCu_{1.7}As_2}$ as well as ${\rm SrCu_2As_2}$ and ${\rm BaCu_2As_2}$ are already in the cT phase at room temperature.

\acknowledgments

This research was supported by the U.S. Department of Energy, Office of Basic Energy Sciences, Division of Materials Sciences and Engineering.  Ames Laboratory is operated for the U.S. Department of Energy by Iowa State University under Contract No.~DE-AC02-07CH11358.

\clearpage

\appendix*
\section{Magnetization Isotherms and Analyses}

\begin{table}
\caption{\label{tab:tableMH} Parameters obtained for ${\rm CaCu_{1.7}As_2}$ from fitting $M(H)$ isotherms at 1.8~K by Eqs.~(\ref{eq:MH_linear-fit}) and (\ref{eq:MH_fit}), where $\theta_{\rm imp} \equiv 0$ and $S_{\rm imp} \equiv 1/2$.  Here $M_{\rm s}$ is the saturation magnetization of ferromagnetic impurities, $\chi$ is the intrinsic susceptibility, and $f_{\rm imp}$ is the molar fraction of paramagnetic impurities.}
\begin{ruledtabular}
\begin{tabular}{lccc}
 field  & $M_{\rm s}$  &  $\chi$   &	$f_{\rm imp}$ \\
 direction & (${\rm G\,cm^3/mol}$) & (${\rm 10^{-5}~cm^3/mol}$) & ($10^{-5}$)\\
\hline
 $H \perp c$      &  0.05(1)  & $-4.27(3)$ & 5.8(1)  \\				
 $H \parallel c$  &  0.00(2)  & $-7.68(7)$ & 7.2(3)   \\		
\end{tabular}
\end{ruledtabular}
\end{table}

\begin{figure}
\includegraphics[width=3in]{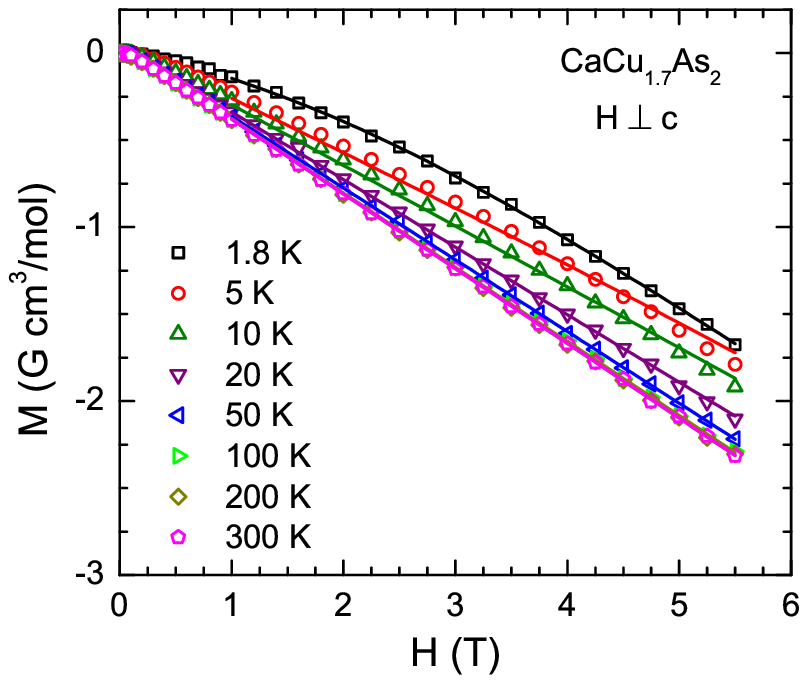}\vspace{0.1in}
\includegraphics[width=3in]{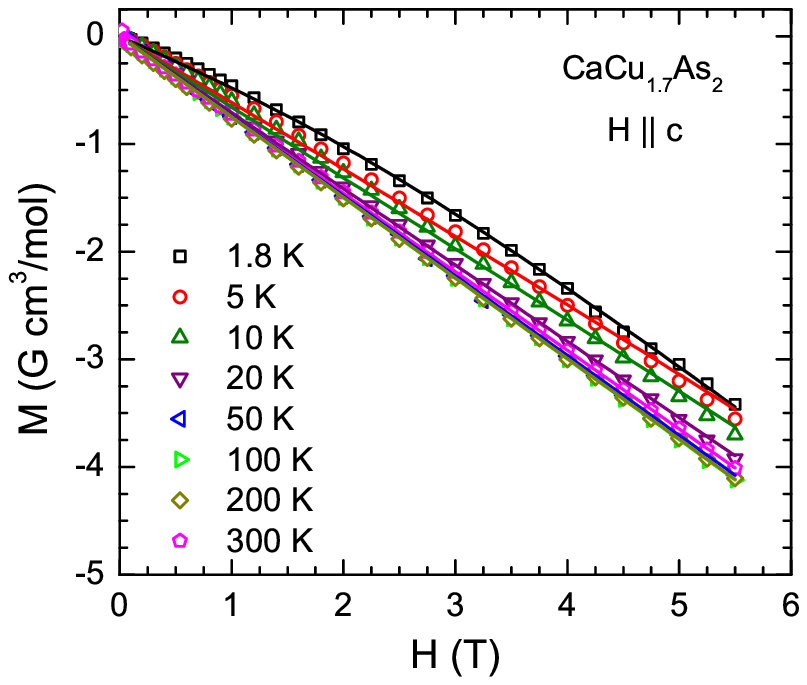}
\caption{(Color online) Isothermal magnetization $M$ of a ${\rm CaCu_{1.7}As_2}$ single crystal as a function of applied magnetic field $H$ measured at the indicated temperatures for $H$ applied (a) in the $ab$-plane ($M_{ab}, H \perp  c$) and, (b) along the $c$-axis ($M_c, H\parallel c$).  The solid curves are fits by Eq.~(\ref{eq:MH_fit}) with $0 \leq H \leq 5.5$~T for $H\parallel c$  and with $1.0 \leq H \leq 5.5$~T for $H \perp c$.}
\label{fig:MH_CaCu2As2}
\end{figure}

Figure~\ref{fig:MH_CaCu2As2} shows $M(H)$ isotherms for a ${\rm CaCu_{1.7}As_2}$ crystal measured at eight temperatures between 1.8 and 300~K with $H$ applied along the $c$-axis ($M_c,\ H \parallel c$) and in the $ab$-plane ($M_{ab},\ H \perp  c$).  The $M(H)$ data exhibit anisotropic diamagnetic behavior with $M_{ab}(H) > M_c(H)$, consistent with the above $\chi(T)$ data.  The $M(H)$ isotherms are almost linear in $H$ for $T \geq 50$~K. For $T \leq 20$~K the $M(H)$ isotherms exhibit slight nonlinearity which we attribute to the presence of saturable paramagnetic impurities in the sample.

In order to estimate the ferromagnetic impurity contribution we fitted the $M(H)$ data at $T\geq50$~K for $H \geq 2$~T by
\begin{equation}
M(H) = M_{\rm s} + \chi H,
\label{eq:MH_linear-fit}
\end{equation}
where $M_{\rm s}$ is the FM impurity saturation magnetization and $\chi$ is the susceptibility. Within the error bars $T$-independent anisotropic values of $M_{\rm s}$ for $H\parallel c$ and $H\perp c$ were obtained as listed in Table~\ref{tab:tableMH}. The $M_{\rm s}$ values indicate that only a trace amount of ferromagnetic impurities is present.  The $M_{\rm s}$ value of 0.05~${\rm G\,cm^3/mol}$ corresponds to the saturation magnetization of 4~molar ppm of Fe metal impurities. However, since ${\rm CaCu_{1.7}As_2}$ is a diamagnetic compound with a small magnitude of $\chi$, even such trace amounts of ferromagnetic impurities are observable in $\chi$ and $M$ measurements.

In order to estimate the paramagnetic impurity contribution to $M$ we fitted the $M(H)$ data for each field direction by
\begin{equation}
M(T,H)=M_{\rm s}+\chi H + f_{\rm imp} M_{\rm{{s_{imp}}}} B_{S_{\rm imp}}(x),
\label{eq:MH_fit}
\end{equation}
where $\chi$ is the intrinsic susceptibility of the compound, $f_{\rm imp}$ is the molar fraction of PM impurities, $M_{\rm {s_{imp}}} = N_{\rm A} g_{\rm imp} \mu_{\rm B} S_{\rm imp}$ is the PM impurity saturation magnetization, $N_{\rm A}$ is Avogadro's number, $\mu_{\rm B}$ is the Bohr magneton, and $g_{\rm imp}$ and $S_{\rm imp}$ are the spectroscopic splitting factor ($g$-factor) and the spin of the impurities, respectively.  Our unconventional definition of the Brillouin function $B_{S_{\rm imp}}$ is given by\cite{Johnston2011}
\begin{subequations}
\label{Eqs:Brillouin}
\begin{eqnarray}
B_{S_{\rm imp}}(x) & =& \frac{1}{2S_{\rm imp}}\Bigg\{\left( 2S_{\rm imp}+1 \right) \coth \left[(2S_{\rm imp}+1)\frac{x}{2} \right]  \nonumber\\
				   && \hspace{2cm} -\ \coth \left( \frac{x}{2} \right)\Bigg\},
\label{eq:Brillouin}
\end{eqnarray}
where
\begin{equation}
x \equiv \frac{g_{\rm imp} \mu_{\rm{B}} H}{k_{\rm{B}} (T-\theta_{\rm imp})}
\end{equation}
\end{subequations}
and we have included a Weiss temperature $\theta_{\rm imp}$ in the argument of $B_{S_{\rm imp}}(x)$ in order to take account of interactions between the paramagnetic impurities in an average mean-field way.  In particular, a Taylor series expansion of Eq.~(\ref{eq:Brillouin}) for $x\ll 1$ yields a Curie-Weiss law for the impurities, {\it i.e.}, $\chi_{\rm imp} = C_{\rm imp}/(T-\theta_{\rm imp})$.

In order to reduce the number of fitting parameters we set the impurity $g$-factor to the fixed value $g_{\rm imp}$ = 2. The $M_{\rm s}^{ab}$ and $M_{\rm s}^{c}$ saturation magnetization values for $H \perp c$ and $H \parallel c$, respectively, were set to the values obtained above from linear fits of the high-field data at $H\geq 2$~T at temperatures $T\geq 50$~K\@.  Since $M_{\rm s}^c$ is zero the $M(H)$ data for $H \parallel c$ were fitted in the whole range $0 \leq H \leq 5.5$~T\@.  However, since $M_{\rm s}^{ab}$ is nonzero the $M(H)$ data for $H \perp c$ were fitted only in the range $1.0 \leq H \leq 5.5$~T\@. The $S_{\rm imp}$ was found to be close to 1/2, so in the final fits we set $S_{\rm imp} = 1/2$. Furthermore, we found that allowing $\theta_{\rm imp}$ to vary during a fit does not make a significant change in the quality of fit and also gives $\theta_{\rm imp}$ close to zero.  Therefore in the final fits we set $\theta_{\rm imp} \equiv  0$ for both field directions.  The final fits of the $M(H)$ data in Fig.~\ref{fig:MH_CaCu2As2} by Eq.~(\ref{eq:MH_fit}) are shown by solid curves in Fig.~\ref{fig:MH_CaCu2As2}.  The fitting parameters obtained for the 1.8~K $M(H)$ isotherms are listed in Table~\ref{tab:tableMH}.  The intrinsic $\chi$ values obtained are plotted in Fig.~\ref{fig:MT_CaCu2As2} as filled blue stars.  These data demonstrate that the intrinsic $\chi$ is independent of $T$ over the measured temperature range $1.8~{\rm K} \leq T \leq 300$~K\@.  We subtracted the ferromagnetic and paramagnetic impurity contributions from the measured $M(T)$ and obtained the intrinsic $\chi$ which is shown by open symbols in Fig.~\ref{fig:MT_CaCu2As2}. The presence of a weak residual upturn in the corrected $\chi \equiv M/H$ versus $T$ data below 25~K indicates that the paramagnetic impurity contribution has not been completely accounted for in these measurements.

\end{document}